\documentclass[10pt,draft]{amsart}
\usepackage{amsmath,amssymb,amsthm}
\begin{document}
\newtheorem{lem}{Lemma}[section]
\newtheorem{prop}{Proposition}[section]
\newtheorem{cor}{Corollary}[section]
\numberwithin{equation}{section}
\newtheorem{thm}{Theorem}[section]

\theoremstyle{remark}
\newtheorem{example}{Example}[section]
\newtheorem*{ack}{Acknowledgments}

\theoremstyle{definition}
\newtheorem{definition}{Definition}[section]

\theoremstyle{remark}
\newtheorem*{notation}{Notation}
\theoremstyle{remark}
\newtheorem{remark}{Remark}[section]

\newenvironment{Abstract}
{\begin{center}\textbf{\footnotesize{Abstract}}%
\end{center} \begin{quote}\begin{footnotesize}}
{\end{footnotesize}\end{quote}\bigskip}
\newenvironment{nome}

{\begin{center}\textbf{{}}%
\end{center} \begin{quote}\end{quote}\bigskip}

\newcommand{\triple}[1]{{|\!|\!|#1|\!|\!|}}

\newcommand{\xx}{\langle x\rangle}
\newcommand{\ep}{\varepsilon}
\newcommand{\al}{\alpha}
\newcommand{\be}{\beta}
\newcommand{\de}{\partial}
\newcommand{\la}{\lambda}
\newcommand{\La}{\Lambda}
\newcommand{\ga}{\gamma}
\newcommand{\del}{\delta}
\newcommand{\Del}{\Delta}
\newcommand{\sig}{\sigma}
\newcommand{\ome}{\omega}
\newcommand{\Ome}{\Omega}
\newcommand{\C}{{\mathbb C}}
\newcommand{\N}{{\mathbb N}}
\newcommand{\Z}{{\mathbb Z}}
\newcommand{\R}{{\mathbb R}}
\newcommand{\Rn}{{\mathbb R}^{n}}
\newcommand{\Rnu}{{\mathbb R}^{n+1}_{+}}
\newcommand{\Cn}{{\mathbb C}^{n}}
\newcommand{\spt}{\,\mathrm{supp}\,}
\newcommand{\Lin}{\mathcal{L}}
\newcommand{\SSS}{\mathcal{S}}
\newcommand{\F}{\mathcal{F}}
\newcommand{\xxi}{\langle\xi\rangle}
\newcommand{\eei}{\langle\eta\rangle}
\newcommand{\xei}{\langle\xi-\eta\rangle}
\newcommand{\yy}{\langle y\rangle}
\newcommand{\dint}{\int\!\!\int}
\newcommand{\hatp}{\widehat\psi}
\renewcommand{\Re}{\;\mathrm{Re}\;}
\renewcommand{\Im}{\;\mathrm{Im}\;}

\title{On the orbital stability for a class of nonautonomous NLS}

\author{Jacopo Bellazzini  $^{1}$ and Nicola Visciglia $^{2}$}
%\thanks{The authors are  partially
%supported by:\\
%M.\ U.\ R.\ S.\ T.\ Prog.\ Nazionale ``Teoria e applicazioni
%delle equazioni iperboliche lineari e non lineari ''.}
 \maketitle

\date{}

\begin{center}
\small $^1$ Dipartimento di Matematica Applicata \\
\small Universit\`a di Pisa\\
\small Via Buonarroti 1/C, 56127 PISA -Italy\\ 
\small $^2$ Dipartimento di Matematica\\
\small Universit\`a di Pisa\\
\small Largo B. Pontecorvo 5, 56127 PISA - Italy\\
\end{center}
\maketitle
\begin{Abstract}
Following the original approach introduced by T. Cazenave
and P.L. Lions in \cite{CaLi}
we prove the existence and the orbital stability
of standing waves for the following class of NLS:
\begin{equation}\label{intr1}
i\partial_t u+ \Delta u - V(x) u + Q(x) u|u|^{p-2}=0,
\hbox{ } (t,x) \in \R\times \R^n, \hbox{ } 2<p<2+\frac 4n
\end{equation}
and
\begin{equation}\label{intr2}
i\partial_t u - \Delta^2 u - V(x) u + Q(x) u|u|^{p-2}=0,
\hbox{ } (t,x) \in \R\times \R^n, \hbox{ } 2<p<2+\frac 8n
\end{equation}
under suitable assumptions on the potentials $V(x)$ and $Q(x)$.
More precisely we assume $V(x), Q(x) \in L^\infty(\R^n)$
and $meas\{Q(x)>\lambda_0\}\in (0,\infty)$ for a suitable $\lambda_0>0$.
The main point is the analysis of the compactness of minimiziang sequences
to suitable constrained minimization problems
related to 
\eqref{intr1} and \eqref{intr2}.
\end{Abstract}
\section{Introduction}

\noindent The aim of this paper is to prove the existence and the orbital stability 
(see Definition \ref{orbital})
of standing waves for a class of Schr\"odinger equations with variable coefficients and with a principal
part that involves both the laplacian operator $\Delta$
and the bilaplacian operator $\Delta^2$.
More precisely the model equations that we consider are of the following type:
\begin{equation}\label{NLSVQ}
i\partial_t u+ \Delta u - V(x) u + Q(x) u|u|^{p-2}=0,
\hbox{ } (t,x) \in \R\times \R^n, \hbox{ } 2<p<2+\frac 4n
\end{equation}
and
\begin{equation}\label{NLSVQbi}
i\partial_t u - \Delta^2 u - V(x) u + Q(x) u|u|^{p-2}=0,
\hbox{ } (t,x) \in \R\times \R^n, \hbox{ } 2<p<2+\frac 8n
\end{equation}
under suitable assumptions on $V(x)$ and $Q(x)$ that will be specified in the sequel. 
We recall that a standing wave solution for \eqref{NLSVQ} (resp.: \eqref{NLSVQbi}), 
is a solution of \eqref{NLSVQ} (resp.: \eqref{NLSVQbi}) of the following type: 
$$u(t, x)\equiv u_0(x)e^{i \omega t} \hbox{ with } u_0\in H^1(\R^n) 
\hbox{ } \hbox{(resp.: } u_0\in H^2(\R^n)\hbox{)}.$$
In particular $u_0(x)$ is solution to one of the following elliptic equations
respectively:
\begin{equation}\label{NLSVQell}
\Delta u_0 - V(x) u_0 + Q(x) u_0|u|_0^{p-2}=\omega u_0
\end{equation}
and 
\begin{equation}\label{NLSVQbiell}
\Delta^2 u_0 + V(x) u_0 - Q(x) u_0|u_0|^{p-2}=-\omega u_0
\end{equation}
for a suitable $\omega\in \R$.
We underline that there is an huge literature devoted to the proof
of the existence of standing waves to \eqref{NLSVQ} and \eqref{NLSVQbi}
(see remarks \ref{caz} and \ref{minc}) under suitable assumptions on $V(x)$ and $Q(x)$. 
However the main point in this paper  
is that, following the general approach introduced in \cite{CaLi}, 
we deduce both the existence and the orbital stability
of standing waves.\\
We recall that in \cite{CaLi}
it is proved the existence and the orbital stability of solitary waves
for a large class of NLS with constant coefficients, 
by analysing the compactness of the minimizing sequences of suitable
minimization problems. 
For instance in \cite{CaLi} the existence and the orbital stability
of standing waves to \eqref{NLSVQ} with $V\equiv 0 $, $Q\equiv 1$
and $2<p<2+\frac 4n$ is deduced
by looking at the following minimization problem:
\begin{equation}\label{inf}
\inf_{{\mathcal M}} 
\left (\frac 12 \int_{\R^n} |\nabla u|^2 dx  - \frac 1p \int_{\R^n} |u|^p dx
\right ), \hbox{ } 2<p<2+\frac 4n\end{equation}
where 
$${\mathcal M}\equiv\left \{u\in H^1(\R^n) | \int_{\R^n} |u|^2 dx=1\right \}.$$
Notice that the energy and the constraint involved in the minimization problem \eqref{inf}
are both quantities preserved along the evolution associated to NLS. More precisely
the following conservation laws occur:
$$E(u(t,x))\equiv \frac 12 \int_{\R^n}  |\nabla u(t,x)|^2 dx - \frac 1p \int_{\R^n} |u(t,x)|^p dx
\equiv E(u(0,x))$$
and
$$\int_{\R^n} |u(t,x)|^2 dx\equiv \int_{\R^n} |u(0,x)|^2 dx$$
where
$u(t,x)$ denotes any finite energy solution to \eqref{NLSVQ} with  $V(x)\equiv 0$
and $Q(x)\equiv 1$. Due to this fact in \cite{CaLi} the authors have been able to deduce
that the set of minimizers to \eqref{inf} are orbitally stables for the flow associated to \eqref{NLSVQ}
with $V(x)\equiv 0$ and $Q(x)\equiv 1$, 
provided that any minimizing sequence for \eqref{inf} is compact in $H^1(\R^n)$ up to the action
of the translations.
Notice that already the proof of the existence of at least a minimizer for \eqref{inf} is not a trivial matter. In fact
the problem is translation invariant and hence
there is an evident lack of compactness. However this difficulty has been overcomed in
\cite{CaLi} by using the concentration--compactness principle (see \cite{Li84}).\\
In this paper, following the general approach introduced in 
\cite{CaLi}, we deduce the existence and the orbital stability
of standing waves to (\ref{NLSVQ}) and (\ref{NLSVQbi}) by minimizing the natural energies
associated to those equations on the constrained manifold
$\|u\|_{L^2}=\rho$ for suitable $\rho>0$. More precisely we consider the following
minimization problems:
\begin{equation}\label{mini1}\inf_{H^1(\R^n)\cap B_{L^2}(\rho)} 
\left (\frac12 \int_{\R^n}|\nabla u|^2 dx + \frac 12 \int_{\R^n} |u|^2 V(x) dx - \frac 1p \int_{\R^n} 
|u|^p Q(x)
dx \right)\end{equation}
and
\begin{equation}\label{mini2}
\inf_{H^2(\R^n)\cap B_{L^2}(\rho)} 
\left (\frac 12 \int_{\R^n}|\Delta u|^2 dx+ \frac 12 \int_{\R^n} |u|^2 V(x) dx - \frac 1p 
\int_{\R^n} |u|^p Q(x)dx
\right)\end{equation}
where
$$B_{L^2}(\rho)\equiv \left \{u\in L^2(\R^n) | \int_{\R^n} |u|^2 dx=\rho^2\right \}.$$
Let us underline that following the same argument as in \cite{CaLi}
one can deduce that the set of minimizers 
to \eqref{mini1} (resp.: \eqref{mini2}) is orbitally stables (see Definition \ref{orbital})
provided that any minimizing sequence to
\eqref{mini1} (resp.: \eqref{mini2}) is compact in $H^1(\R^n)$ (resp.: $H^2(\R^n))$.
We skip the proof of this implication, since it 
is essentially contained in \cite{CaLi}.
Hence, since now on we shall be mainly 
concentrated on the analysis of the minimizing sequences to \eqref{mini1} and \eqref{mini2}.\\
Notice that the main difficulty related to the minimization problems 
\eqref{mini1} and \eqref{mini2}
is connected with
the lack of compactness of the Sobolev embedding $H^s(\R^n)$
in the corresponding $L^q(\R^n)$ spaces. Moreover, due to the
presence of the coefficients $V(x)$ and $Q(x)$ in \eqref{mini1} 
and \eqref{mini2}, our minimization problems are not invariant
by the action of the translations, despite to the problems studied  in \cite{CaLi}.
In order to overcome those difficulties we state below an abstract
variational principle that guarantees the compactness of minimizing sequences for 
a general family of minimization problems.\\
In the sequel we need the following 
\begin{definition}\label{weakly}
Let $\mathcal H$ be an Hilbert space and $T:{\mathcal H} \rightarrow \R$ a nonlinear map.
We say that $T$ is {\em weakly continuous } if
for every sequence $h_k \in \mathcal H$ the following implication is satisfied:
$$h_{k}\rightharpoonup \bar h \hbox{ in } \mathcal H
\Rightarrow
 \lim_{k\rightarrow \infty} T(h_{k})=T(\bar h).$$
\end{definition}

\begin{prop}\label{liebg}
Let $({\mathcal H}, \|.\|)$,$({\mathcal H}_1, \|.\|_1)$, $({\mathcal H}_2, \|.\|_2)$
be three Hilbert spaces such that:$${\mathcal H}\subset {\mathcal H}_1,
\hbox{ } {\mathcal H}\subset {\mathcal H}_2$$
and 
\begin{equation*}
c (\|h\|_{{\mathcal H}_1}^2+\|h\|_{{\mathcal H}_2}^2)\leq \|h\|_{\mathcal H}^2\leq C (\|h\|_{{\mathcal H}_1}^2 +\|h\|_{{\mathcal H}_2}^2)  
\hbox{ } \forall h\in \mathcal H. 
\end{equation*}
Assume also that are given $\rho>0$ and two functionals
$S,T:{\mathcal H}\rightarrow \R$
such that:
\begin{itemize}
\item \begin{equation}\label{(1)}T(0)= 0;
\end{equation}
\item \begin{equation}\label{(2)}
\hbox{ $T$ is weakly continuous;}
\end{equation}
\item 
\begin{equation}\label{(3)}
T(\lambda h)\leq \lambda^2 T(h) \hbox{
and  } S(\lambda h)\leq \lambda^2 S(h) \hbox{ }\forall \lambda\geq 1, h \in \mathcal H;\end{equation}
\item 
\begin{equation}\label{(4)}
\hbox{ if } h_k\rightharpoonup \bar h \hbox{ in } {\mathcal H}
\hbox{ and } h_k\rightarrow \bar h \hbox{ in } {\mathcal H}_2
\end{equation}
$$\hbox{ then } S(h_k)\rightarrow S(\bar h);$$
\item  
\begin{equation}\label{(5)}\hbox{ if } h_k\rightharpoonup \bar h
\hbox{ in } \mathcal H \hbox{ then } 
S(h_k-\bar h) + S(\bar h)=S(h_k)+ o(1);\end{equation}
\item \begin{equation}\label{(6)}-\infty<I_{S,T}^\rho<I_{S}^\rho\end{equation}
where
\begin{equation}\label{prturb}
I_{S,T}^\rho\equiv \inf_{h\in B_{{\mathcal H}_2}(\rho)\cap {\mathcal H}} 
\left (\frac 12\|h\|^2_{\mathcal H_1} + S(h) + T(h)\right )
\end{equation}
$$I_{S}^\rho\equiv \inf_{h\in B_{{\mathcal H}_2}(\rho)\cap {\mathcal H}} \left 
(\frac 12 \|h\|^2_{\mathcal H_1} + S(h)\right )$$
and $$B_{{\mathcal H}_2}(\rho)\equiv \{h\in {\mathcal H}_2 | \|h\|_{{\mathcal H}_2}=\rho\};$$
\item \begin{equation}\label{(7)}
{\hbox{for every sequence } \{h_k\}\in B_{{\mathcal H}_2}(\rho)\cap \mathcal H
\hbox{ such that }}
\|h_k\|_{\mathcal H}\rightarrow \infty, \end{equation}$$\hbox{ we have }
\frac 12 \|h_k\|_{{\mathcal H}_1}^2 + S(h_k)+ T(h_k)\rightarrow \infty \hbox{ as }
k\rightarrow 	\infty.$$
\end{itemize}
\noindent Then every minimizing sequence for \eqref{prturb}, 
i.e. 
$$h_k\in B_{{\mathcal H}_2}(\rho)\cap \mathcal H  \hbox{ and }
\frac 12 \|h_k\|^2_{{\mathcal H}_1} + S(h_k) + T(h_k)
\rightarrow I_{S,T}^\rho,$$
is compact in ${\mathcal H}$.
\end{prop}

\begin{remark}\label{concom}
There are many concrete situations in which
all the assumptions of Proposition \ref{liebg}
are satisfied (see for example the proof of Theorems
\ref{NLS} and \ref{NLSbi}).
Next we show shortly how
the existence of a minimizer for \eqref{inf} can be deduced by using Proposition
\ref{liebg}.
In fact due to a symmetrization argument it is sufficient
to restrict \eqref{inf} on the space $H^1_{rad} (\R^n)$
of the radially symmetric functions belonging to $H^1(\R^n)$.
Next we make the following concrete choice
of the spaces: ${\mathcal H}\equiv H^1_{rad} (\R^n)$,
 ${\mathcal H}_1\equiv {\mathcal D}^{1,2}_{rad} (\R^n)$ and ${\mathcal H}_2\equiv L^2_{rad} (\R^n)$,
where ${\mathcal D}^{1,2}_{rad} (\R^n)$ and $L^{2}_{rad} (\R^n)$
denote the radially symmetric functions belonging to the classical Beppo Levi space 
${\mathcal D}^{1,2}(\R^n)$ and to the Lebesgue space $L^2(\R^n)$ respectively.
Finally as functionals $S$ and $T$ we choose
respectively $S\equiv 0$ (i.e. the trivial operator)
and $T(u)=-\frac 1p \int_{\R^n} |u|^p dx$.
It is easy to check that in this specific framework 
all the hypothesis
of Proposition \ref{liebg} are satisfied. In particular it is easy to check
\eqref{(1)}, \eqref{(3)}, \eqref{(4)}, \eqref{(5)} while condition \eqref{(2)}
follows from the 
compactness of the embedding 
$H^1_{rad} (\R^n)\subset L^p(\R^n)$.
By combining an elementary convexity argument with the Sobolev embedding
it is easy to show that the l.h.s inequality 
in \eqref{(6)} and \eqref{(7)} are satisfied.
Finally notice that the r.h.s. inequality in \eqref{(6)} in this specific framework
becomes $I<0$, where $I$ is the infimum defined in \eqref{inf}.
The  proof of this inequality follows from an elementary rescaling argument.

\end{remark}

\noindent In order to state our main results 
it is necessary to give the precise  definition of orbital stability
(see \cite{CaLi}) that we state below for completeness.
First we introduce for every $V(x), Q(x)\in L^\infty(\R^n)$ the following quantities:
\begin{equation}\label{B21}
I_{\rho,p}^{V,Q}
\end{equation}\begin{equation*}
=\inf_{H^1(\R^n)\cap B_{L^2}(\rho)} 
\left (\frac12 \int_{\R^n}|\nabla u|^2 dx + \frac 12 \int_{\R^n} |u|^2 V(x) dx - \frac 1p \int_{\R^n} |u|^p Q(x)
dx \right)\end{equation*}
and
\begin{equation}\label{B22}J_{\rho,p}^{V, Q}\end{equation}
\begin{equation*}=
\inf_{H^2(\R^n)\cap B_{L^2}(\rho)} 
\left (\frac 12 \int_{\R^n}|\Delta u|^2 dx + \frac 12 \int_{\R^n} |u|^2 V(x) dx - \frac 1p \int_{\R^n} |u|^p Q(x)dx
\right)\end{equation*}
where
$$B_{L^2}(\rho)\equiv \left \{u\in L^2(\R^n) | \int_{\R^n} |u|^2 dx=\rho^2\right \}.$$

\begin{definition}\label{orbital}
The set of minimizers $${\mathcal M}_{\rho, p}^{V,Q}\equiv \left\{u\in H^1(\R^n)
\cap B_{L^2}(\rho)|
u \hbox{ is minimizer for }  
I_{\rho,p}^{V,Q} \right\}$$
(resp.: 
${\mathcal N}_{\rho, p}^{V,Q}\equiv \left\{u\in H^2(\R^n)\cap B_{L^2}(\rho) | u \hbox { is any minimizer for }  
J_{\rho,p}^{V,Q} \right\})$\\ 
is {\it orbitally stable} 
for \eqref{NLSVQ} (resp.: \eqref{NLSVQbi}) 
provided that:
\begin{equation*}
\forall \varepsilon, \ \exists \delta>0 \ \text{s.t.} \  \forall
v \in H^1(\R^N) \hbox{(resp.: $H^2(\R^n))$}
\end{equation*}
\begin{equation*}
\hbox{ with }\inf_{u\in {\mathcal M}_{\rho, p}^{V,Q}} \|v-u\|_{H^1(\R^n)}<\delta 
\hbox{ (resp.: $\inf_{u\in {\mathcal M}_{\rho, p}^{V,Q}} \|v-u\|_{H^2(\R^n)}<\delta$)}\
\text{, then }
\end{equation*}
\begin{equation*}
\forall t\geq 0 \ \ \inf_{u\in {\mathcal M}_{\rho, p}^{V,Q}}
\|v(t,\cdot)-u\|_{H^1(\R^n)}<\varepsilon
\hbox{ (resp.: $\inf_{u\in {\mathcal N}_{\rho, p}^{V,Q}}
\|v(t,\cdot)-u\|_{H^2(\R^n)}<\varepsilon$)}
\end{equation*}
where $v(t,x)$ is the solution of \eqref{NLSVQ} (resp. \eqref{NLSVQbi}) 
with initial data $v(x)$.
\end{definition}
\noindent As a consequence of Proposition \ref{liebg} we prove 
Theorem \ref{NLS} that mainly concerns the compactness of any minimizing sequence
to \eqref{mini1} under suitable assumptions on $V(x)$ and $Q(x)$.
In this way, by using the general argument 
in \cite{CaLi} already mentioned above, we get the orbital stability
of the set ${\mathcal M}_{\rho, p}^{V,Q}$ introduced in Definition \ref{orbital}. 
In the sequel we denote by $meas (A)$ the Lebesgue measure of the
measurable set $A$.
\begin{thm}\label{NLS}
Let $2<p<2 + \frac 4n$ and $V(x), Q(x)\in L^\infty(\R^n)$.
Assume that:
$$Q(x)\geq 0 \hbox{ a.e. }  x\in \R^n;$$
\begin{equation}\label{hyp}
\hbox{ there is } \lambda_0>0 \hbox{ s.t. } 0<meas \{Q(x)>\lambda_0\}<\infty.
\end{equation}
Then there exists $\rho_0>0$ such that
all the minimizing sequences for \eqref{B21}
are compact in $H^1(\R^n)$ provided that $\rho>\rho_0$.
In particular ${\mathcal M}_{\rho,p}^{V,Q}$
is a no empty compact set and it is orbitally stable.
\end{thm}
\begin{remark}\label{caz}
Let us underline that 
the existence of standing waves for 
\eqref{NLSVQ} has been extensively studied
in the literature. We mention in particular \cite{Bo},
\cite{DingNi}, \cite{Li84}, \cite{WZ}
where existence results are proved under quite general assumptions
on $V(x)$, $Q(x)$ and under the following stronger assumption
$2<p<\frac{2n}{n-2}$. However the main point in Theorem \ref{NLS} 
is that we prove the existence
of standing waves by looking at the minimization problem \eqref{B21}
and hence we also deduce, following \cite{CaLi}, their orbital stability.
\end{remark}
\begin{remark}
Notice that, since the solutions constructed in Theorem \ref{NLS} are orbitally stables, it is quite 
natural to consider nonlinearities that grow 
at the rate $2<p<2+\frac 4n$. In fact the value $p=2+\frac 4n$ 
is the critical power for which the
Cauchy problem associated to the focusing NLS is globally well--posed (for more details on this point see
the comments in \cite{CaLi}).
\end{remark}
\begin{remark}
Concerning the assumption
\eqref{hyp}, notice that it is satisfied 
for instance by any function $Q(x)\in {\mathcal C}(\R^n)$ such that
$$\sup_{\R^n} Q(x)>\limsup_{|x|\rightarrow \infty} Q(x),$$ which is 
a condition very much exploited in \cite{DingNi}
and \cite{WZ}. However we underline that in Theorem \ref{NLS} a larger class of 
functions $Q(x)$ is allowed.
\end{remark}
\begin{remark} 
We point out that, following the approach in \cite{PrVi} and \cite{PrVi2},
in Theorem \ref{NLS} are allowed functions $V(x)$ and $Q(x)$ 
which are not necessarily continuous. As far as we know this is another novelty
in this paper compared with previous existence results.  
\end{remark}
\noindent Next we state a version of Theorem \ref{NLS}
for equation \eqref{NLSVQbi}.
Recall that the set ${\mathcal N}_{\rho,p}^{V,Q}$ has been introduced in Definition
\ref{orbital}.
\begin{thm}\label{NLSbi}
Let $2<p<2+\frac 8n$ and $V(x), Q(x)\in L^\infty(\R^n)$.
Assume that:
$$Q(x)\geq 0 \hbox{ a.e. } x\in \R^n;$$
\begin{equation}\label{hypbi}
\hbox{ there is } \lambda_0>0 \hbox{ s.t. } 0<meas \{Q(x)>\lambda_0\}<\infty.
%m=\inf_{\R^n} V(x)< 
%\liminf_{x\rightarrow \infty} V(x)=m_\infty
\end{equation} 
Then there exists $\rho_0>0$ 
such that
all the minimizing sequences for \eqref{B22}
are compact in $H^2(\R^n)$, provided that $\rho>\rho_0$.
In particular ${\mathcal N}_{\rho,p}^{V,Q}$ 
is a no empty compact set and it is orbitally stable.
\end{thm}
\begin{remark}\label{minc}
Concerning previous existence results of standing waves for \eqref{NLSVQbi}
let us mention \cite{Li84} where the 
existence of solutions is proved via the concentration--compactness argument 
by assuming that the nonlinear term is regular and has a suitable asymptotic behaviour.  
In \cite{Be1} equation \eqref{NLSVQbiell} is treated in the specific case $V(x) \equiv 1$
and with an autonomous nonlinearity $f(u)$.
In this case the existence of solutions is proved by means of
the compact embedding 
of $H^2_{rad}(\R^n)$ in  $L^p(\R^n)$ for $2<p<\frac{2n}{n-4}$. 
Many efforts have been devoted to study elliptic equations
involving the biharmonic operator with critical nonlinearity
$f(u)\equiv u|u|^{\frac 8{n-4}}$,  see for instance \cite{Al} and \cite{Cha}.
Finally let us mention \cite{De1} where a deep analysis of the bifurcation properties of the operator 
$\Delta^2  -\lambda$ is done.\\
In the best of our knowledge the existence result stated in Theorem \ref{NLSbi}
under such a general assumptions on $V(x)$ and $Q(x)$ is not written elsewhere.
Moreover,
since we deduce the existence result in Theorem
\ref{NLSbi} by looking at the minimization problem \eqref{B22}  
we also get, following \cite{CaLi}, the orbital stability of the corresponding solutions. As far as we know the 
question of the stability
of solitary waves to \eqref{NLSbi} has not been analysed in previous papers.    
\end{remark}

\vspace{0.1cm}

\noindent Next we fix some notations.
For every $x\in \R^n, R>0$ we denote by $B_R(x)$ the ball in $\R^n$ of radius
$R$ and centered in $x$.\\
For every $1\leq p\leq \infty$ we denote by $\|.\|_p$ the classical
$L^p(\R^n)$--norm and
for every measurable set $A$ we denote by $meas (A)$ its Lebesgue measure.\\
Given $\rho>0$ we use the notation
$$B_{L^2}(\rho)\equiv \left \{u\in L^2(\R^n) | \int_{\R^n} |u|^2 dx=\rho^2\right \}.$$

\vspace{0.1cm}

\noindent The paper is organized ad follows: section 2 is devoted to the proof of Proposition \ref{liebg},
in sections 3 and 4 we prove respectively Theorem
\ref{NLS} and \ref{NLSbi}. In the Appendix 
we show the existence of a minimizer for \eqref{mini2} when $V(x)\equiv 0$
and $Q(x)\equiv 1$.
This fact is important along the proof of Theorem
\ref{NLSbi}.
Let us point out that the content of the Appendix  
is essentially contained in \cite{Li84},
however for the sake of completeness we give all the details of the proof. 

\section{Proof  of Proposition \ref{liebg}}\label{abstract}
 
\noindent By assumption \eqref{(7)} we have that $\sup_{k\in \N}\|h_k\|_{\mathcal H}<\infty$
hence we can assume $h_k \rightharpoonup \bar h$ in $\mathcal H$.
On the other hand we have
\begin{equation}\label{comp12g}\|h_k\|_{{\mathcal H}_2}=\rho 
\hbox{ and } \frac 12 \|h_k\|^2_{{\mathcal H}_1} + S(h_k)+T(h_k)=I_{S,T}^\rho+ o(1)
\end{equation}
and due to assumption \eqref{(2)} we get
\begin{equation}\label{compg}T(h_k)= T(\bar h) + o(1).\end{equation}
By combining \eqref{comp12g} and \eqref{compg} we get:
\begin{equation}\label{newtar}
\frac 12 \|h_k\|^2_{{\mathcal H}_1}+ S(h_k)+T(\bar h) =I_{S,T}^\rho + o(1).
\end{equation}
In particular this implies
$$I_S^\rho+T(\bar h)\leq I_{S,T}^\rho$$
and hence due to \eqref{(1)} and \eqref{(6)} we get $\bar h\neq 0$.\\ 
On the other hand by combining \eqref{newtar} with \eqref{(5)} we get: 
\begin{equation}\label{newtar2}
\frac 12 \|h_k-\bar h\|^2_{{\mathcal H}_1} + \frac 12
\| \bar h\|^2_{{\mathcal H}_1}+ 
S(h_k-\bar h) + S(\bar h)\end{equation}$$+T(\bar h)=I_{S,T}^\rho+ o(1).$$
Moreover we have
\begin{equation}\label{bellaz}
\|h_k - \bar h\|_{{\mathcal H}_2}^2
= \|h_k\|_{{\mathcal H}_2}^2 - 
\|\bar h\|_{{\mathcal H}_2}^2+ o(1)
= \rho^2- \|\bar h\|_{{\mathcal H}_2}^2+o(1)\end{equation}
and hence, since $\bar h\neq 0$, we deduce that
\begin{equation}\label{13gio}
\|h_k - \bar h\|_{{\mathcal H}_2}^2< \rho^2
\end{equation}
provided that $k$ is large enough.
Next notice that since $\bar h$ is the weak limit of the sequence
$\{h_k\}$ and since
$\|h_k\|_{{\mathcal H}_2}=\rho$, we get 
\begin{equation}\label{12g}
\|\bar h\|_{{\mathcal H}_2}\leq \rho.\end{equation} 
By combining \eqref{newtar2}, \eqref{13gio}, \eqref{12g} with the  
assumption \eqref{(3)} we deduce
\begin{equation}\label{1g}
\frac{\|h_k-\bar h\|_{{\mathcal H}_2}^2}{\rho^2}
\left ( \frac 12 \left  \| \frac{\rho(h_k-\bar h)}
{\|h_k-\bar h\|_{{\mathcal H}_2}}\right \|_{{\mathcal H}_1}^2 
+   
S\left ( \frac{\rho(h_k-\bar h)}{\|h_k-\bar h\|_{{\mathcal H}_2}}\right)
\right )
\end{equation}
\begin{equation*}
+\frac{\|\bar h\|_{{\mathcal H}_2}^2}{\rho^2}
\left ( \frac 12 \left \|\frac{\rho \bar h}{\|\bar h\|_{{\mathcal H}_2}}
\right \|_{{\mathcal H}_1}^2 
+ 
T\left ( \frac{\rho \bar h}{\|\bar h\|_{{\mathcal H}_2}}\right)+ 
S\left ( \frac{\rho \bar h}{\|\bar h\|_{{\mathcal H}_2}}\right)\right )
\end{equation*}$$\leq I_{S,T}^\rho +o(1)
$$
and hence due to \eqref{bellaz}
\begin{equation}\label{23g}
\frac 1{\rho^2}(\|h_k-\bar h\|_{{\mathcal H}_2}^2
I_S^\rho + 
\|\bar h\|_{{\mathcal H}_2}^2 I_{S,T}^\rho)
\end{equation}$$\leq I_{S,T}^\rho\frac 1{\rho^2}(\|\bar h\|_{{\mathcal H}_2}^2 
+\|h_k-\bar h\|_{{\mathcal H}_2}^2)+o(1),$$
that in turn implies:
$$\|h_k-\bar h\|_{{\mathcal H}_2}^2
(I_S^\rho- I_{S,T}^\rho)\leq o(1).$$
By \eqref{(6)} we get
$h_k\rightarrow \bar h$ strongly in 
${\mathcal H}_2$.
It is now easy to deduce by \eqref{(4)} that
the convergence occurs strongly also in $\mathcal H$.

\hfill$\Box$

\section{Applications to NLS}

\noindent Given $\rho>0$, $p>2$ and $U(x),W(x)\in L^\infty(\R^n)$ 
we define the following quantities:
$$I_{\rho,p}^{U,W}=\inf_{H^1(\R^n)\cap B_{L^2}(\rho)} 
\left (\frac 12\|\nabla u\|_2^2 + \frac 12\int_{\R^n} |u|^2 U(x) dx - \frac 1p \int_{\R^n} |u|^p W(x)
dx \right).$$
Moreover if $V(x)\in L^\infty(\R^n)$ is as in Theorem \ref{NLS},
then we introduce the new function  
\begin{equation}\label{newV}
\tilde V(x)\equiv V(x) - \hbox{{\em infess}}_{\R^n} V(x).
\end{equation}
Notice that $\tilde V$ is a non--negative function, i.e.
$\tilde V(x)\geq 0 \hbox{ a.e. } x\in \R^n$. This property will be important
along  the proof of next Lemma.
\begin{lem}\label{condition}
Assume that $V(x)$, $Q(x)$, $\lambda_0$ and $p$ 
are as in Theorem \ref{NLS}. Then there exists 
$\rho_0>0$ such that:
$$I_{\rho,p}^{\tilde V,Q}<I_{\rho,p}^{\tilde V, Min \{Q, \lambda_0\}}
\hbox{ } \forall \rho>\rho_0.$$
\end{lem}

\noindent{\bf Proof.} Since $Q(x)\in L^\infty(\R^n)$ 
we deduce by the Lebesgue derivation Theorem that:
$$\lim_{\delta \rightarrow 0} \delta^{-n}\int_{B_\delta(x_0)} |Q(x)-Q(x_0)| \hbox{ } dx=0
\hbox{ a.e. }  x_0\in {\R}^n.$$
Since by assumption we have
$meas \{x\in {\R}^n|Q(x)> \lambda_0\}>0,$
we deduce that there exists $\bar x\in \{x\in {\R}^n|Q(x)> \lambda_0\}$
such that 
$\lim_{\delta \rightarrow 0} \delta^{-n}\int_{B_\delta(\bar x)} 
|Q(x)-Q(\bar x)| \hbox{ } dx=0.$
For simplicity we can assume that $\bar x\equiv 0$ and 
also $\bar \lambda\equiv Q(0)$, hence we have
\begin{equation}\label{lebesgue}
\lim_{\delta \rightarrow 0} \delta^{-n}\int_{B_\delta(0)} |Q(x)
- \bar \lambda| \hbox{ } dx=0 \hbox{ with } \bar \lambda > \lambda_0.
\end{equation}
Next we fix a minimizer $u_0\in H^1(\R^n)$ 
for 
$I_{1,p}^{0, \lambda_0}$ (the proof of the existence of 
a minimizer is given in \cite{Li84}, see also
remark \ref{concom}).
Then it is easy to check
that
\begin{equation}\label{rescalingV}
u_0^\rho\equiv u_0\left(\frac x{\rho^\alpha}\right)\rho^{-\beta}
\hbox{ is a minimizer for }
I_{\rho,p}^{0, \lambda_0},\end{equation}
where
\begin{equation}\label{alphabetaV}
\alpha=\alpha(n, p)=\frac{2(p-2)}{(p-2)n-4} \hbox{ and }
\beta=\beta(n,p)=\frac 4{(p-2)n-4}.
\end{equation}
We claim that there is $\rho_0>0$ such that
\begin{equation}\label{claimV}I_{\rho,p}^{\tilde V,Q}
<I_{\rho,p}^{0, \lambda_0}\hbox{ } \forall \rho>\rho_0.
\end{equation}
On the other hand $\tilde V(x)\geq 0$ and $0\leq Min\{Q(x),\lambda_0\}\leq \lambda_0$ 
and this implies 
\begin{equation}\label{abbV}
I_{\rho,p}^{0,\lambda_0}\leq I_{\rho,p}^{\tilde V, Min \{Q(x), \lambda_0\}}.
\end{equation}
By combining \eqref{claimV} with \eqref{abbV}
we deduce the desired result.

\noindent Next we prove \eqref{claimV}. Due
to \eqref{rescalingV} it is sufficient to prove the following inequality:  
$$\frac 12\|\nabla u_0^\rho\|_2^2 + \frac 12\int_{\R^n} |u_0^\rho |^2 \tilde V(x)
dx - \frac 1p \int_{\R^n} |u_0^\rho|^p Q(x)dx 
$$$$<\frac 12 \|\nabla u_0^\rho \|_2^2 
- \frac {\lambda_0}{p} \int_{\R^n} |u_0^\rho|^p dx$$
or equivalently 
\begin{equation}\label{equivV}
I(\rho) + II(\rho)
+III(\rho)
\end{equation}
\begin{equation*}
\equiv \frac 12 \int_{\R^n} |u_0^\rho|^2 \tilde V(x) dx
-\frac 1p \int_{\R^n} |u_0^\rho|^p (Q(x)-\bar \lambda)dx 
+ \frac{\lambda_0 - \bar \lambda}{p}
\int_{\R^n} |u_0^\rho|^p dx<0.
\end{equation*}
Let us fix $R_0>0$ such that
\begin{equation}\label{Ro}
\frac 2p\|Q\|_\infty\int_{|x|>R_0} |u_0|^p dx
+\frac{\lambda_0 - \bar \lambda}{p}\|u_0\|_p^p=-\epsilon_0<0,
\end{equation}
and notice that due to \eqref{lebesgue} we get:
$$I(\rho) \leq \frac 12 \|\tilde V\|_\infty \|u_0^\rho\|_2^2 = \frac 12 \|\tilde V\|_\infty\rho^2;$$
$$II(\rho)\leq \frac 1p \rho^{-p\beta}\|u_0\|_\infty^p
\int_{B(0,R_0\rho^\alpha)} |Q(x)-\bar \lambda| dx
+ \frac 2p\|Q\|_\infty \int_{|x|>R_0\rho^\alpha} |u_0^\rho|^p dx 
$$
$$= o(1) \rho^{n\alpha-p\beta} + \frac 2p\|Q\|_\infty\rho^{(n\alpha -p\beta)}
\int_{|x|>R_0} |u_0|^p dx
$$
where $\lim_{\rho\rightarrow \infty} o(1)=0$;
$$III(\rho)=\frac{(\lambda_0 - \bar \lambda)}p
\int_{\R^n}|u_0^\rho|^p dx=\frac{(\lambda_0 - \bar \lambda)}p
\rho^{n\alpha-p\beta}\|u_0\|_p^p.$$
Hence we finally get
\begin{equation}\label{claimprimeV}
I(\rho) + II(\rho)+ III(\rho)
\end{equation}
$$\leq \frac 12 \|\tilde V\|_\infty \rho^2 - \epsilon_0
\rho^{n\alpha-p\beta} \|u_0\|_p^p + o(1)\rho^{n\alpha-p\beta} $$
where $\epsilon_0>0$ is the constant that appear in \eqref{Ro}.
Since
$2< n\alpha - p \beta$, the inequality \eqref{claimprimeV} implies
\eqref{equivV} for $\rho$ large enough,
and in turn it is equivalent to
\eqref{claimV}.

\hfill$\Box$

\noindent{\bf Proof of Theorem \ref{NLS}}
Notice that by definition of $\tilde V$ (see \eqref{newV}) we have
$V-\tilde V\equiv const$. Due to this fact it is easy to deduce that
the minimizing sequences for
$I^{V, Q}_{\rho,p}$ are precisely the minimizing sequences for
$I^{\tilde V, Q}_{\rho,p}$. Hence it is sufficient to show
that any minimizing sequence for  $I^{\tilde V, Q}_{\rho,p}$ is compact in $H^1(\R^n)$
in order to deduce the same property for the minimizing sequences 
for $I^{V, Q}_{\rho,p}$.
In order to prove this fact we use Proposition \ref{liebg}
where we make the following specific choice for the spaces 
${\mathcal H}, {\mathcal H}_1,{\mathcal H}_2$ and for the operators $S$ and $T$:
$${\mathcal H}= H^1(\R^n), \mathcal{H}_1= {\mathcal D}^{1,2}(\R^n),
\mathcal{H}_2=L^2(\R^n),$$
$$S(u)=\int_{\R^n} \left 
(\frac 12 |u|^2 \tilde V(x) - \frac 1p |u|^p \min
\{Q(x), \lambda_0\}\right ) dx, 
$$$$T(u)=-\frac 1p \int_{Q(x)\geq \lambda_0} |u|^p 
(Q(x)-\lambda_0) dx$$
and we also assume $\rho>\rho_0$
where $\rho_0$ is the same constant that appears in Lemma
\ref{condition}. It is easy to verify that in this specific framework
the assumptions $\eqref{(1)}, \eqref{(3)}$ in Proposition \ref{liebg}
are satisfied. 
The r.h.s inequality in \eqref{(6)} follows from
Lemma \ref{condition} provided that $\rho>\rho_0$,
\eqref{(2)} is a consequence of
the Rellich Compactness Theorem in conjunction with hypothesis \eqref{hyp}.
Condition \eqref{(4)} comes from the following elementary
fact:
$$\hbox{ if } u_k\rightarrow  \bar u \hbox{ in } L^2(\R^n) \hbox{ and } 
u_k \hbox{ is bounded in } H^1(\R^n)$$$$ \hbox{ then }
u_k\rightarrow  \bar u \hbox{ in } L^p(\R^n) \hbox{ } \forall 
\hbox{ } 2\leq p <\frac{2n}{n-2}.$$
In order to deduce the l.h.s. inequality \eqref{(6)} and \eqref{(7)},
notice that by combining the convexity inequality
with the Sobolev embedding we get:
$$\|u\|_{p}\leq C \|\nabla u\|_{2}^{\frac{n(p-2)}{2p}}\|u\|_{2}^{1-\frac n2 +\frac np}
= C \|\nabla u\|_{2}^{\frac{n(p-2)}{2p}}\rho^{1-\frac n2 +\frac np}
\hbox{ }\forall u\in H^1(\R^n)\cap B_{L^2}(\rho)$$
and this implies
\begin{equation}\label{crucpal}
\frac 12 \|\nabla u\|_2^2 +\int_{\R^n} |u|^2 \tilde V(x) dx - \int_{\R^n} 
|u|^p Q(x) dx \end{equation}
$$\geq \frac 12 \|\nabla u\|_2^2 - C(\rho) \|Q\|_{\infty} \|\nabla u\|_2^{\frac{n(p-2)}{2}} 
$$
where we have used $\tilde V(x) \geq 0$.
Since we are assuming $2<p<2+\frac 4n$ we have 
$2>\frac{n(p-2)}{2}$ and hence by \eqref{crucpal} 
we can deduce both the l.h.s. inequality in \eqref{(6)} and \eqref{(7)}.
Finally we shall check condition \eqref{(5)}.
From the Rellich Compactness Theorem we get:
\begin{equation}\label{rellich}
\hbox{ if } u_k\rightharpoonup  \bar u \hbox{ in } H^1(\R^n)
\hbox{ then } u_k\rightarrow \bar u \hbox{ in } L^p(B(0, R)) \hbox{ } \forall R>0.
\end{equation}
It is also easy to check that:
$$\hbox{ if } u_k\rightharpoonup  \bar u \hbox{ in } H^1(\R^n)
\hbox{ then }$$$$ \int_{\R^n} |u_k|^2 \tilde V(x) dx= 
\int_{\R^n} |u_k-\bar u|^2 \tilde V(x) dx +\int_{\R^n} |\bar u|^2 \tilde V(x) dx 
+ o(1)$$
hence \eqref{(5)}
will follow from:
\begin{equation}\label{equiv(5)}
\lim_{k\rightarrow \infty} \left |\int_{\R^n} (|u_k|^p  
-  |u_k-\bar u|^p 
- |\bar u|^p) \min \{Q(x), \lambda_0\} dx \right|=0
\end{equation}
provided that  $u_k\rightharpoonup  \bar u \hbox{ in } H^1(\R^n)$.
Since $\bar u \in L^p(\R^n)$ we have:
\begin{equation}\label{infin}
\lim_{R\rightarrow \infty} \int_{\R^n \setminus B(0, R)} |\bar u|^p \min \{Q(x), \lambda_0\} dx=0.
\end{equation}
On the other hand since $p>2$ 
we can use the mean value theorem 
in order to deduce the existence of a constant $C=C(p)>0$ such that
\begin{equation}
\left ||t+h|^p - |t|^p\right | 
\leq C |h| (|t|+ |h|)^{p-1}
\hbox{ } \forall t, h \in \R.
\end{equation}
In particular for every $k\in \N$ we have
$$\left ||u_k(x)-\bar u(x)|^p 
- |u_k(x)|^p\right | \leq C |\bar u(x)| 
(|u_k(x)|+ |\bar u(x)|)^{p-1} \hbox{ a.e. }  x\in \R^n$$
that by the H\"older inequality implies 
$$\int_{\Omega} ||u_k(x)-\bar u(x)|^p 
- |u_k(x)||^p \min \{Q(x), \lambda_0\}
dx$$$$\leq C \|\min \{Q(x), \lambda_0\}\|_{L^\infty(\R^n)}
\|\bar u\|_{L^p(\Omega)}
\left (\||u_k| + |\bar u| \|_{L^p(\Omega)}^{\frac{p-1}p}\right ) $$
for every measurable set $\Omega\subset \R^n$.
By combining this estimate (where we choose $\Omega=\R^n \setminus B(0, R)$)
with the uniform boundedness of
$\|u_k\|_{L^p(\R^n)}$ (that follows from
the usual Sobolev embedding $H^1(\R^n) \subset L^p(\R^n)$) 
and  with \eqref{infin} we get
\begin{equation}\label{uniformity}
\forall \epsilon>0 \hbox{ } \exists R(\epsilon)>0 \hbox{ s.t. } |r(k, \epsilon)|<\epsilon
\hbox{ } \forall k\in \N \hbox{ where }\end{equation}
\begin{equation*}
r(k,\epsilon)\equiv \int_{\R^n \setminus B(0,R(\epsilon))} (|u_k-\bar u|^p 
- |u_k|^p) \min \{Q(x), \lambda_0\} dx. 
\end{equation*}
Moreover due to \eqref{rellich} we get
\begin{equation*}
\lim_{k\rightarrow \infty} s(k, \epsilon)=0 \hbox{ and } \lim_{k\rightarrow \infty} t(k, \epsilon)=0
\hbox{ }
\forall \epsilon>0
\end{equation*}
where 
$$s(k, \epsilon)\equiv \int_{B(0,R(\epsilon))}
(|u_k|^p -|\bar u|^p) \min \{Q(x), \lambda_0\} dx$$
and
$$t(k, \epsilon)\equiv \int_{B(0,R(\epsilon))}
|u_k -\bar u|^p \min \{Q(x), \lambda_0\} dx.$$
In particular
we deduce that there exists $k(\epsilon)\in \N$ such that
\begin{equation}\label{corol12}
|s(k,\epsilon)| \hbox{ and } |t(k,\epsilon)|<\epsilon \hbox{ } \forall k>k(\epsilon).
\end{equation}
Moreover we have
$$\int_{\R^n} |u_k|^p \min \{Q(x), \lambda_0\} dx $$$$=
\int_{\R^n \setminus B(0,R(\epsilon))} |u_k|^p \min \{Q(x), \lambda_0\} dx + 
\int_{B(0,R(\epsilon))}|u_k|^p \min \{Q(x), \lambda_0\} dx$$
$$=-r(k,\epsilon)+ \int_{\R^n \setminus B(0,R(\epsilon))} |u_k-\bar u|^p \min \{Q(x), \lambda_0\} dx 
$$$$+s(k,\epsilon)+ 
\int_{B(0,R(\epsilon))}|\bar u|^p \min \{Q(x), \lambda_0\} dx$$
$$=-r(k,\epsilon)- t(k,\epsilon)+ \int_{\R^n} |u_k-\bar u|^p \min \{Q(x), \lambda_0\} dx 
$$$$+s(k,\epsilon)+ \int_{\R^n}|\bar u|^p \min \{Q(x), \lambda_0\} dx
-\int_{\R^n \setminus B(0,R(\epsilon))}|\bar u|^p \min \{Q(x), \lambda_0\} dx.$$
Moreover \eqref{uniformity} and \eqref{corol12}
imply
$$|r(k, \epsilon)| + |s(k, \epsilon)| + |t(k, \epsilon)|<3\epsilon \hbox{ } \forall k>k(\epsilon)$$
and hence
$$\left |\int_{\R^n} (|u_k|^p  
-  |u_k-\bar u|^p 
- |\bar u|^p) \min \{Q(x), \lambda_0\} dx \right|
$$$$<3\epsilon + \int_{\R^n \setminus B(0,R(\epsilon))}|\bar u|^p \min \{Q(x), \lambda_0\} dx
\hbox{ } \forall k>k(\epsilon).$$
Due to \eqref{infin} and since $R(\epsilon)\rightarrow \infty$ as $\epsilon\rightarrow 0$
it is now easy to deduce \eqref{equiv(5)}.

\hfill$\Box$

\section{Applications to NLS involving the biharmonic operator}

\noindent The proof of Theorem \ref{NLSbi} follows the same lines
of the proof of Theorem \ref{NLS}.
A basic tool is the following version of Lemma
\ref{condition} where, given $\rho>0$, $p>2$ and $U(x),W(x)\in L^\infty(\R^n)$, the quantity 
$J_{\rho,p}^{U,W}$ is defined as follows:
$$J_{\rho,p}^{U,W}=\inf_{H^2(\R^n)\cap B_{L^2}(\rho)} 
\left (\frac 12\|\Delta u\|_2^2 + \frac 12\int_{\R^n} |u|^2 U(x) dx - \frac 1p \int_{\R^n} |u|^p W(x)
dx \right).$$
In next Lemma the potential $\tilde V(x)$ is the one introduced in \eqref{newV}.
\begin{lem}\label{conditionbi}
Assume that $V(x)$, $Q(x)$, $\lambda_0$ and $p$ 
are as in Theorem \ref{NLSbi}. Then there exists $\rho_0>0$ such that
$$J_{\rho,p}^{\tilde V,Q}<J_{\rho,p}^{\tilde V,\min\{Q(x), \lambda_0\}}
\hbox{ } \forall \rho>\rho_0.$$
\end{lem}

\noindent It is easy to check that the proof of Lemma
\ref{condition} can be adapted in order to prove Lemma \ref{conditionbi},
provided that we are able to show the existence of a minimizer
to the following minimization problem:
$$J_{\rho,p}^{0,\lambda_0}=
\inf_{H^2(\R^n)\cap B_{L^2}(\rho)} 
\left (\frac 12 \|\Delta u\|_2^2 - \frac{\lambda_0}p \int_{\R^n} |u|^p dx
\right).$$
The detailed proof of this fact 
is given for completeness in the Appendix, even if it is essentially contained in 
\cite{Li84}.
We skip the detailed proof of Lemma \ref{conditionbi}.
 
\vspace{0.1cm}

\noindent{\bf Proof of Theorem \ref{NLSbi}}
As in the proof of Theorem \ref{NLS}, it sufficient to prove the compactness
of the minimizing sequences for $J^{\tilde V, Q}_{\rho,p}$.
This fact can be shown by using Proposition \ref{liebg} in the following explicit framework:
$${\mathcal H}= H^2(\R^n), \mathcal{H}_1= {\mathcal D}^{2,2}(\R^n),
\mathcal{H}_2=L^2(\R^n),$$
$$S(u)=\int \left (\frac 12|u|^2 \tilde V(x)  - \frac 1p |u|^p \min
\{Q(x), \lambda_0\}\right ) dx, 
$$$$T(u)=-\frac 1p \int_{Q(x)\geq \lambda_0} |u|^p 
 (Q(x)-\lambda_0)) dx$$
and $\rho>\rho_0$ where $\rho_0$ is the constant that appears in Lemma
\ref{conditionbi}. It is easy to check, following the same arguments
involved in the proof of Theorem \ref{NLS}, that all the assumptions
required in Proposition \ref{liebg} are satisfied in this specific context.
In particular the r.h.s inequality in \eqref{(6)} follows from Lemma \ref{conditionbi}.
However we skip the details of the proof.

\hfill$\Box$

\section{Appendix}

\noindent The aim of this Appendix is to give a detailed proof of 
Theorem \ref{minNLSapp}. We recall that it is essentially contained in \cite{Li84} however,
since it plays a fundamental role along the proof of Theorem \ref{NLSbi},
we give all the details of the proof for completeness. \\
\noindent
In the sequel we denote by $Q_R(y)$ the cube 
$$\left [y_1-\frac R2, y_1 +\frac R2\right )\times...\times 
\left [y_n-\frac R2, y_n +\frac R2\right )$$
where $y\equiv (y_1,...., y_n)\in \R^n$ and $R>0$.
 
\begin{thm}\label{minNLSapp}
For every $\rho>0$ and for every $2<p<2+\frac 8n$, there is a minimizer
for the following problem:
\begin{equation}\label{Jrop}
J_{\rho,p}=\inf_{H^2(\R^n)\cap B_{L^2}(\rho)} 
\left (\frac 12 \|\Delta u\|_2^2 - \frac 1p \int_{\R^n} |u|^pdx \right).
\end{equation}
\end{thm}
\begin{remark}
In fact we prove a stronger version of Theorem \ref{minNLSapp}.
More precisely we show 
that all the minimizing sequences for $J_{\rho,p}$ are compact in $H^2(\R^n)$ up to the action of the
translations. 
\end{remark}
\noindent In the sequel we need the following
\begin{prop}\label{boundedness}
If $2<p<2+\frac{8}{n}$ and $\rho>0$, then:
\begin{itemize}
\item $J_{\rho,p}>-\infty$;
\item any
minimizing sequence $u_k\in H^2(\R^n)\cap B_{L^2}(\rho)$ for \eqref{Jrop}
is bounded in $H^2(\R^n)$;
\item $J_{\rho,p}<0.$
\end{itemize}
\end{prop}
\noindent{\bf Proof}
By combining the Sobolev embedding $\|u\|_{\frac{2n}{n-4}}\leq C \|\Delta u\|_{2}$
with the convexity inequality we get:
\begin{equation}
\|u\|_{p}\leq C \|u\|_{2}^{\frac{2n-np+4p}{4p}}\|\Delta u\|_{2}^{\frac{n(p-2)}{4p}}.
\end{equation}
From this inequality it is easy to deduce that the functional 
$$H^2(\R^n)\cap B_{L^2}(\rho) \ni u \rightarrow \frac 12 \|\Delta u\|_2^2 - \frac 1p \int_{\R^n} |u|^pdx,$$
is bounded from below  and moreover every minimizing sequence is bounded in $H^2(\R^n)$,
provided that  $2<p<2+\frac{8}{n}$.\\
Finally we shall prove that $J_{\rho,p}<0$. 
We fix $v\in H^2(\R^n)\cap B_{L^2}(\rho)$, then 
$v_{\lambda}\equiv \lambda^{-\frac{n}{2}}v(\frac{x}{\lambda})$ 
belongs to $H^2(\R^n)\cap B_{L^2}(\rho)$. Moreover we have
\begin{equation*}
J_{\rho,p}
\leq \frac 12 \|\Delta v_{\lambda}
\|_{2}^2-\frac 1p\int_{\R^n} |v_{\lambda}|^p dx 
$$$$
=\frac{1}{2\lambda^4} \int_{\R^n} |\Delta v|^2 dx - \frac{\lambda^{n(1-\frac{p}{2})}}p
\int_{\R^n} |v|^p dx \hbox{ } \forall \lambda>0
\end{equation*}
and this implies $J_{\rho,p}<0$ provided that we choose $\lambda$ in a suitable way.

\hfill$\Box$

\begin{prop}\label{fundvanish}
Let $u_k$ is a sequence bounded in $H^2(\R^n)$ and such that
\begin{equation}\label{nonvani}
\lim_{k\rightarrow \infty} \left (\sup_{y\in \R^n} \int_{Q_1(y)} |u_k|^2 dx \right )
=0.
\end{equation}
Then $u_k\rightarrow 0$ in $L^{q}(\R^n)$ for any $q \in \left (2, \frac{2N}{N-4}\right )$.
\end{prop}
\noindent {\bf Proof}
By the convexity inequality we get
$$\|u_k\|_{L^{\frac{2(n+8)}{n+4}}(Q_1(y))}^\frac{2(n+8)}{n+4}
\leq \|u_k\|_{L^2(Q_1(y))}^\frac 8{n+4} \|u_k\|_{L^{2+\frac 8n}(Q_1(y))}^2 \hbox{ } \forall k\in \N, y\in \R^n.$$
Next we introduce for every $z\equiv (z_1,..., z_n)\in \Z^n$
the cube
$$Q_z\equiv [z_1, z_1+1)\times....\times[z_n, z_n+1).$$
Notice that $\R^n$ is the disjoint union of the cubes $Q_z$ with $z\in \Z^n$.
Hence we get
$$\|u_k\|_{\frac{2(n+8)}{n+4}}^\frac{2(n+8)}{n+4}
=\sum_{z\in \Z}\|u_k\|_{L^{\frac{2(n+8)}{n+4}}(Q_z)}^\frac{2(n+8)}{n+4}
$$$$\leq \sum_{z\in \Z^n} \|u_k\|_{L^2(Q_z)}^\frac 8{n+4} \|u\|_{L^{2+\frac 8n}(Q_z)}^2 
\leq C \left (\sup_{y\in \R^n} \int_{Q_1(y)} |u_k|^2 dx \right)^\frac 4{n+4}
\sum_{z\in \Z}\|u_k\|_{H^2(Q_z)}^2$$
$$=\left (\sup_{y\in \R^n} \int_{Q_1(y)} |u_k|^2 dx \right)^\frac 4{n+4} \|u_k\|_{H^2(\R^n)}^2$$
where we have the Sobolev embedding
$H^2(Q_z)\subset L^{2+\frac 8n}(Q_z).$
By using the assumptions on $u_k$ we finally deduce:
$$\lim_{k\rightarrow \infty}\|u_k\|_{\frac{2(n+8)}{n+4}}=0.$$
By combining this fact, with the embedding
$$H^2(\R^n)\subset L^2(\R^n)\cap L^\frac{2n}{n-4}(\R^n),$$
and with the convexity inequality we get
$$\lim_{k\rightarrow \infty}\|u_k\|_{q}=0 \hbox{ } \forall q\in \left (2, \frac{2n}{n-4}\right ).$$

\hfill$\Box$

\begin{prop}\label{subadditivity}
For any $\rho>0$ and $0<\mu<\rho$ the following subadditivity condition holds
\begin{equation*}
J_{\rho,p}<J_{\mu,p}+J_{\sqrt{\rho^2-\mu^2},p}.
\end{equation*}
\end{prop}
\noindent{\bf Proof}
It is sufficient to prove \begin{equation*}
J_{\theta \rho,p}< \theta^2 J_{\rho,p} \hbox{ } \forall \rho>0, \theta>1.
\end{equation*}
We take $u\in H^2(\R^n)\cap B_{L^2}(\rho)$ and we set $u_\theta\equiv \theta u(x)$. Clearly $\|u_\theta \|_{2}^2=\theta^2\rho^2$ and thus
\begin{eqnarray}
 J_{\theta\rho,p}\leq \inf_{H^2(\R^n)\cap B_{L^2}(\rho)}
 \left (\frac {\theta^2} 2 \|\Delta u\|_2^2 - \frac{\theta^p} p \int_{\R^n} |u|^p dx \right )\\
 <  \inf_{H^2(\R^n)\cap B_{L^2}(\rho)} \theta^2 \left (\frac 1 2  \|\Delta u\|_2^2dx - \frac 1p  \int_{\R^n} |u|^pdx \right) = \theta^2 J_{ \rho,p}.
\end{eqnarray}
\hfill$\Box$

\noindent{\bf Proof of Theorem \ref{minNLSapp}}
Let $u_{k}\in H^2(\R^n)\cap B_{L^2}(\R^n)$ be a minimising sequence for $J_{\rho,p}$, then 
due to Proposition \ref{boundedness} we have that 
up to a subsequence we can assume $u_k \rightharpoonup \bar u$ in $H^2(\R^n)$. Moreover 
$\bar u$ satisfies one of the following  conditions:
\begin{itemize}
\item $\|\bar u\|_{2}=0$;\\
\item $0 <\|\bar u\|_{2}<\rho$;\\
\item $\|\bar u\|_{2}=\rho.$
\end{itemize}
Notice also that for any sequence $y_k\in \R^n$ we have that
$u_k(.+y_k)$ is still a minimizing sequence for $J_{\rho,p}$.
This implies that the proof of the Theorem can be concluded 
provided that we show the existence of a sequence $y_k\in \R^n$ such that
the weak limit of $u_k(.+y_k)$ belongs to $B_{L^2}(\rho)$.
 
\vspace{0.1cm} 

{\em First step:  there exists a sequence $y_k\in \R^n$ s.t. $u_k(.+y_k) \rightharpoonup \bar v \hbox{ and } 
0<\|\bar v\|_2\leq \rho$}

\vspace{0.1cm}

\noindent It is sufficient to show that 
\begin{equation}\label{nelben}
\sup_{y\in \R^n} \int_{Q_1(y)} |u_k|^2 dx \geq \mu>0.
\end{equation}
In fact in this case
we can fix $y_k\in \R^n$ such that
$$\int_{Q_1(0)} |u_k (.+y_k)|^2 dx\geq \mu>0$$
and hence, due to the compactness of the embedding $H^2(Q_1(0))\subset L^2(Q_1(0))$,
we deduce that the weak limit of the sequence $u_k (.+y_k)$ is not the trivial function.
It is then sufficient to show \eqref{nelben}. If by the absurd
\eqref{nelben} is false then by 
Proposition \ref{fundvanish} we get: 
$$
\lim_{k\rightarrow \infty} \int_{\R^n} |u_k|^p dx = 0
$$
and hence  
$$J_{\rho,p}=\lim_{k\rightarrow \infty} \left (\frac 12 \|\Delta u_k\|_2^2 
-\frac 1p \int_{\R^n} |u_k|^p dx\right )= \lim_{k\rightarrow \infty} \frac 12 \|\Delta u_k\|_2^2 \geq 0$$
which is in contradiction with the property $J_{\rho,p}<0$ (see Proposition
\ref{boundedness}). 

\vspace{0.1cm} 

{\em Second step : if $\bar v$ is as in the first step, then $\|\bar v\|_2=\rho$.}

\vspace{0.1cm}

\noindent For simplicity we use the notation
$$v_k\equiv u_k(.+y_k)$$ 
where $y_k\in \R^n$ is the sequence given in the first step.
Notice also that, due to the first step, we have  
$$\|\bar v\|^2_{2}=\theta^2\in (0, \rho^2].$$
Next assume by the absurd that $0<\theta^2<\rho^2$.
Notice that we have:
$$\rho^2=\|v_k - \bar v\|^2_{2}
+ \|\bar v\|_2^2+o(1)
\hbox{ i.e. }
\|v_k - \bar v\|^2_{2}=\rho^2-\theta^2 + o(1);$$
$$\|\Delta v_k\|_2^2= \|\Delta (v_k - \bar v)\|^2_{2}
+ \|\Delta \bar v\|_2^2+o(1);$$
$$\|v_k-\bar v\|_p^p+ \|\bar v\|_p^p= \|v_k\|_p^p + o(1)
\hbox{ (see the proof of \eqref{equiv(5)})}.$$
Hence by combining all those facts we deduce:
$$J_{\theta,p} + J_{\sqrt{\rho^2-\theta^2},p}
%\leq J_p\left (\frac{\theta \bar v}{\|\bar v\|_2}\right )+ J_p\left (\frac{\sqrt{\rho^2-\theta^2}(v_k-\bar %v)}{\|v_k-\bar v\|_2}\right )
$$$$\leq \frac 12 \|\Delta \bar v\|_2^2 - \frac 1p \|\bar v\|_p^p
+ \frac 12 \frac{\rho^2 - \theta^2}{\|v_k-\bar v\|_2^2}\|\Delta \bar v - \Delta v_k\|_2^2 
-\frac 1p \|v_k-\bar v\|_p^p \frac{(\rho^2-\theta^2)^\frac p2}{\|v_k-\bar v\|_2^p} $$ 
$$= \frac 12  \|\Delta \bar v\|_2^2 - \frac 1p \|\bar v\|_p^p
+ \frac 12 \|\Delta \bar v - \Delta v_k\|_2^2 
-\frac 1p \|v_k-\bar v\|_p^p 
+o(1)$$ 
$$= \frac 12 \|\Delta v_k\|_2^2 - \frac 1p \|v_k\|_p^p
+o(1)\rightarrow J_{\rho,p}$$
and this is in contradiction with Proposition \ref{subadditivity}.
Hence necessarily we have $\theta^2=\rho^2$.

\hfill$\Box$


\begin{thebibliography}{99}
\bibitem{Al}Alves, M. J.; Carrião, P. C.; Miyagaki, O. H.
{\it Soliton solutions to a class of quasilinear elliptic equations on $\Bbb R$},
Adv. Nonlinear Stud. {\bf 7}, no. 4, 579--597 (2007) 

%\bibitem{BBGM}  Bellazzini J., Benci V., Ghimenti M., 
%Micheletti A., {\it On the existence of the fundamental eigenvalue
%of an elliptic problem in $\R^n$},
% Adv. Nonl. Studies {\bf 7}, 551--564 (2007).

\bibitem{Be1} Bellazzini J., {\it A nonlinear eigenvalue problem for the biharmonic
operator in $\R^N$}, Preprint (2007)

\bibitem{Bo} Bonheure D., J. Van Schaftingen, {\it Bound state solutions for a class of nonlinear Schr\"odinger equations}, Rev. Mat. Iberoamericana {\bf 24}, no. 1, 297--351 (2008).

%\bibitem{BL83} Brezis, H.; Lieb, E., {\it A relation between pointwise convergence of functions and convergence of %functionals.}  Proc. Amer. Math. Soc.  {\bf 88}, no. 3, 486--490 (1983) 

\bibitem{CaLi} Cazenave T., Lions P.L.,
{\it Orbital Stability of Standing Waves for Some Non linear
Schr\"odinger Equations}, Commun. Math. Phys. {\bf 85}, 549--561
(1982)

\bibitem{Cha}  Chabrowski J., do O M., {\it On some fourth-order semilinear elliptic problems in $R^N$}, Nonlinear Analysis TMA { \bf 49} ,  861--884   (2002)



\bibitem{De1} Deng Y., Li Y., {\it 
Branches of solutions to semilinear biharmonic equations on 
$\Bbb R\sp N$ },
Proc. Roy. Soc. Edinburgh Sect. A {\bf136}, 733--758 (2006) 


\bibitem{DingNi} Ding W.Y., Ni W.M. {\it On the existence of 
positive entire solutions of a semilinear elliptic equation},
Arch. Rat. Math. Anal.
{\bf 31}, 283--308 (1986).

\bibitem{Li84}  Lions P.L.,
{\it The concentration-compactness principle in the Calculus of
Variations. The locally compact case, part 2}, Ann. Inst. Henri
Poincar\'e {\bf 1}, 223--283(1984)

\bibitem{PrVi} Prinari F., Visciglia N.
{\it On a minimization problem involving the critical Sobolev exponent},
Adv. Nonl. Studies {\bf 7},  551--564 (2007).


\bibitem{PrVi2} Prinari F., Visciglia N. {\it 
Standing waves for a class of Schr\"odinger equations with potentials in
$L^\infty$}, to appear on Hokkaido Mathematics Journal.

\bibitem{WZ} Wang X., Zeng B. {\it On concentration of positive bound states of
Nonlinear Schr\"odinger Equations with competing potential functions},
Siam J. Math. Anal. {\bf 28}, no. 3, 633--655 (1997).

\end{thebibliography}
\end{document}